\documentclass[11pt]{article}

\usepackage{amsmath}
\usepackage{graphicx}
\usepackage{indentfirst}
\usepackage{amssymb}
\usepackage{cite}
\usepackage{color}
\usepackage{subfigure}

\newcommand{\rc}{r_{\rm c}}
\newcommand{\ri}{r_{\rm i}}
\newcommand{\ro}{r_{\rm o}}

\newcommand{\Lam}{\Lambda}
\newcommand{\rcir}{r_{\rm cir}}
\newcommand{\riro}{\ri=\ro}
\newcommand{\rorc}{\ro=\rc}

\begin{document}

\title{\bf{Charged Particle and Strong Cosmic Censorship in Reissner--Nordstr\"{o}m--de Sitter Black Holes}}

\date{}
\maketitle

\begin{center}
\author{Yongwan Gim}$^{a}$\footnote{yongwan89@sogang.ac.kr} and \author{Bogeun Gwak}$^{b}$\footnote{rasenis@dgu.ac.kr}

\vskip 0.25in
$^{a}$\it{Department of Physics and Research Institute for Basic Science, Sogang University, Seoul 04107, \\ Republic of Korea}\\
$^{b}$\it{Division of Physics and Semiconductor Science, Dongguk University, Seoul 04620,\\Republic of Korea}\\

\end{center}
\vskip 0.6in

{\abstract
{We investigate the instability of the unstable circular orbit of a charged null particle to test the strong cosmic censorship conjecture in Nariai-type near-extremal Reissner--Nordstr\"{o}m--de Sitter black holes. The instability is estimated as the Lyapunov exponent and found to depend on the mass and charge of the black hole. Then, we explicitly show that charged null particles in unstable circular orbits correspond to the charged massless scalar field in the eikonal limit. This provides a compact relationship representing the quasinormal frequency in terms of the characteristics of unstable circular orbits in near Nariai-type extremal conditions. According to this relationship, the strong cosmic censorship conjecture is valid.
}}

\thispagestyle{empty}
\newpage
\setcounter{page}{1}

\section{Introduction}\label{sec1}

Black holes are intriguing astronomical objects. The event horizon covers the inside of a black hole; according to classical mechanics, none of the massless and massive particles can escape from the gravity of a black hole by crossing the event horizon. In other words, classical black holes only absorb matter. However, black holes can lose energy through Hawking radiation through a quantum effect\cite{Hawking:1974sw,Hawking:1976de}. Thus, black holes are considered as thermodynamic systems with the Hawking temperature expressed in terms of surface gravity, and the Bekenstein--Hawking entropy expressed in terms of the area of the black hole \cite{Bekenstein:1973ur,Bekenstein:1974ax}. Recently, the existence of black holes was experimentally proven by the Laser Interferometer Gravitational-Wave Observatory (LIGO), which detected the signals of gravitational waves from the collisions between black holes.

At the center of some black holes there exists a curvature singularity enclosed by the event horizon or the Cauchy horizon. Without horizons, the naked singularity results in the failure of the predictability of the Einstein equations. Hence, all singularities, except for the Big Bang singularity, should be hidden inside the horizons of black holes such that they are invisible to the observers; this is the cosmic censorship conjecture\cite{Penrose:1964wq, Penrose:1969pc, Hawking:1969sw}. There are two kinds of conjectures---weak conjecture and strong conjecture. The weak cosmic censorship (WCC) conjecture indicates that the singularities inside black holes should be hidden from distant observers such that the event horizon is stable against perturbations. The WCC conjecture was first investigated in a rotating black hole, where the addition of a particle cannot overspin the black hole beyond the extremal condition\cite{Wald:1974ge}. As there is no general theorem to prove the validity of the WCC conjecture, it has been tested in various black holes through particle absorption\cite{Hubeny:1998ga, Crisostomo:2003xz, Jacobson:2009kt, Barausse:2010ka,Isoyama:2011ea,Barausse:2011vx, Gao:2012ca,Hod:2013vj,Rocha:2014jma,Gwak:2015fsa,Cardoso:2015xtj, Colleoni:2015ena, Colleoni:2015afa, Horowitz:2016ezu, Gwak:2017kkt,Sorce:2017dst, Revelar:2017sem,Duztas:2017lxk,Ge:2017vun,Yu:2018eqq,Shaymatov:2018fmp,Gim:2018axz,Zeng:2019jrh} and scattering of the test field\cite{Hod:2008zza,Semiz:2005gs,Toth:2011ab,Duztas:2013wua,Semiz:2015pna,Natario:2016bay,Gwak:2018akg, Duztas:2018adf,Chen:2018yah,Gwak:2019asi,Chen:2019nsr,Gwak:2019rcz}.

A strong cosmic censorship (SCC) conjecture suggests that the singularities inside black holes should be space-like \cite{Penrose:1969pc, Hawking:1970}; a historical review of the SCC conjecture can be found in \cite{Chambers:1997ef}. In reality, in-falling observers can identify space-like singularities, but these cannot escape from inside black holes owing to the space-like geometry. However, if the black hole possesses an inner horizon, namely, the Cauchy horizon, the singularity inside the black hole becomes a timelike singularity. This leads to the failure of the SCC conjecture. Therefore, the instability of the Cauchy horizon is an important issue in the context of the SCC conjecture. For an asymptotic flat black hole such as Reissner--Nordstr\"{o}m black holes and the Kerr black holes, the inner horizon is unstable if the horizon is perturbed by external fields; the perturbation field is infinitely blue-shifted at the Cauchy horizon. Thus, the singularity is in the space-like geometry, which proves the validity of the SCC conjecture \cite{Matzner:1979zz,Poisson:1990eh,Ori:1991zz,Brady:1995ni}. However, for Reissner--Nordstr\"{o}m-de Sitter (RNdS) black holes, the perturbation field can be red-shifted at the Cauchy horizon owing to the existence of a cosmological horizon. Nevertheless, it has been suggested that relevant modes can provide the instability of the inner horizon; therefore, RNdS black holes should not be considered a counterexample to the SCC conjecture\cite{Brady:1998au}. Recently, \cite{Cardoso:2017soq} claimed that the decay rates of neutral perturbation fields are determined by the quasinormal modes (QNMs) of RNdS black holes, and that the SCC conjecture is violated in near-extremal RNdS black holes. Since then, the validity of the SCC conjecture in RNdS black holes has been controversial \cite{Luna:2018jfk,Ge:2018vjq,Destounis:2018qnb,Rahman:2018oso, Gwak:2018rba}. In particular, this argument has been extended to the charged perturbation fields in RNdS black holes \cite{Burikham:2017gdm,Hod:2018dpx,Cardoso:2018nvb,Dias:2018etb,Mo:2018nnu, Dias:2018ufh,Liu:2019rbq,Zhang:2019ynp}. Recently, the validity of the SCC conjecture was investigated using circular null geodesics of a neutral particle in the near-extremal Kerr--Newman--de Sitter black hole \cite{Hod:2018lmi}. Furthermore, the SCC conjecture is now extended and investigated to various black holes\cite{Liu:2019lon,Gwak:2019ttv,Guo:2019tjy,Gan:2019jac,Destounis:2019omd}.

Interestingly, the unstable circular orbits of a null particle were recently studied under the SCC conjecture. The unstable circular orbits of the neutral null particles are directly related to the QNMs of neutral scalar fields \cite{Mashhoon:1985cya, Cardoso:2008bp}. The scalar field with the QNMs satisfies the boundary conditions of purely ingoing waves at the event horizon and purely outgoing waves at the cosmological horizon of de Sitter black holes. For the neutral scalar field, the real and imaginary parts of the QMNs in the eikonal limit are given as a compact relationship with the angular velocity and the Lyapunov exponent of the null particle in the unstable circular geodesics\cite{Cardoso:2008bp}. The Lyapunov exponent determines the stability of the particles along the circular null geodesics; the particle starts to leak out of the geodesics at this instability time scale, which is an inverse of the Lyapunov exponent \cite{Cardoso:2008bp}. The imaginary part of the QNMs determines the characteristic time scale for the decay of the neutral scalar fields. According to the compact relationship in \cite{Cardoso:2008bp}, the Lyapunov exponent of a neutral massless particle in the null circular geodesic is connected with the imaginary part of the QNM of neutral scalar fields in the eikonal limit. Therefore, one can understand the instability of the QNMs of the neutral scalar fields in the eikonal limit in terms of the instability of the charged particles leaking out of the unstable circular null orbits.

In this work, we investigate the instability of the unstable circular null trajectory of the {\it charged null particle} around RNdS black holes. Then, the Lyapunov exponent is obtained and related to the QNMs of the {\it charged massless scalar field} in the eikonal limit to understand and discuss the SCC conjecture. Here, we find a {\it compact relationship} between the charged particles and scalar field in RNdS black holes under the near Nariai-type extremal condition. The compact relationship of the charged case is first obtained in our analysis. The compact relationship for the {\it neutral particle} for RNdS black holes was discussed in \cite{Cardoso:2008bp}, for higher-dimensional RNdS black holes in\cite{Rahman:2018oso}, for Kerr--dS black holes in\cite{Dias:2018ynt}, for higher-dimensional Kerr--dS black holes in\cite{Rahman:2018oso}, and for Kerr--Newman--dS black holes in\cite{Hod:2018lmi}. Therefore, our compact relationship is the generalization to that of the neutral cases. Particularly, the compact relationship can be only obtained under the near Nariai-type extremal condition, and so it has a very rare analytical form relating the Lyapunov exponent and quasinormal frequency of the charged null particle and charged scalar field. Furthermore, we find that black holes have small critical exponents that indicate the instability of the circular null trajectory of the charged null particle. According to the critical exponent, the instability of the unstable circular trajectory depends on the ratio between the mass and charge of the black hole. Finally, the validity of the SCC conjecture in Nariai-type near-extremal RNdS black holes is investigated through the Lyapunov exponent of the circular orbit of the charged null particles.

The remainder of this paper is organized as follows: In Sec.\,\ref{sec2}, we introduce the geometry of the RNdS black holes. In Sec.\,\ref{sec3}, we derive the geodesic motion of the charged particle from the Hamilton--Jacobi equations in the RNdS black hole, and subsequently, we define the effective potential for the radial motion of a charged particle in the RNdS black holes. In Sec.\,\ref{sec4}, we calculate the instability time scale, typical orbital time scale, and the critical exponent of the circular null orbit of the charged null particle. In Sec.\,\ref{sec5}, we show that the compact relation is valid even between the charged null particle and the charged scalar field in the RNdS black hole with the near Nariai-type extremal condition. In Sec.\,\ref{sec6}, using the result of Sec.\,\ref{sec5}, we investigate the validity of the SCC conjecture in the case of the Nariai-type near-extremal RNdS black hole. Finally, the study is summarized in Sec.\,\ref{sec7}.

\section{Reissner--Nordstr\"om--de Sitter Black Holes}\label{sec2}

RNdS black holes are the solution to the four-dimensional Einstein--Maxwell gravity with a positive cosmological constant. The action is 
\begin{align}\label{eq:EMCaction1}
{\cal S}_{\rm G}=\frac{1}{16\pi}\int d^4 x \sqrt{-g} \left(R-F_{\mu\nu}F^{\mu\nu}-2\Lambda\right),
\end{align}
where $R$, $F_{\mu\nu}$, and $\Lambda$ denote the curvature, Maxwell field strength, and cosmological constant, respectively. The Maxwell field strength is given by a gauge field $A_\mu$ with an electric charge $Q$ as
\begin{align}
F_{\mu\nu}=\partial_\mu A_\nu - \partial_\nu A_{\mu}, \quad A=-\frac{Q}{r}dt.
\end{align}
An RNdS black hole is a static solution. The metric with the mass $M$ and charge $Q$ is 
\begin{align}\label{eq:metric02}
ds^2 = - \frac{\Delta(r)}{r^2}dt^2+ \frac{r^2}{\Delta(r)}dr^2+r^2 d\theta^2 +r^2 \sin^2\theta d\phi^2,\quad \Delta(r)=-\frac{\Lambda r^4}{3}+r^2-2Mr+Q^2.
\end{align}
Since the RNdS black hole is asymptotically the dS spacetime, there are three horizons: a cosmological horizon $\rc$, an outer horizon $\ro$, and an inner horizon $\ri$. Their radial locations range as $\ri<\ro<\rc$. In this regard, the metric function $\Delta$ in Eq.\,\eqref{eq:metric02} can be rewritten in terms of these horizons as
\begin{equation}\label{eq:metric03}
\Delta(r)=\frac{\Lambda}{3}(r_{\rm c}-r)(r-r_{\rm o})(r-r_{\rm i})(r+\rc+\ro+\ri).
\end{equation}
Comparing the metric functions in Eq.\,\eqref{eq:metric02} with Eq.\,\eqref{eq:metric03}, we can rewrite the cosmological constant $\Lam$, mass $M$, and electric charge $Q$ of the RNdS black hole in terms of the horizons as
\begin{align}
\Lam =\frac{3}{\rc^2 + \ro^2 + (\rc+\ri)(\ro + \ri)},\,M = \frac{1}{2}\frac{(\rc +\ro) (\rc + \ri) (\ro + \ri)}{ \rc^2 + \ro^2 + (\rc +\ri) (\ro + \ri)},\,Q^2 =\frac{\rc\ri\ro(\rc+\ro+\ri)}{\rc^2 + \ro^2 + (\rc+\ri)(\ro + \ri)}.
  \label{eq:MQL}
\end{align}
Surface gravities are associated with the amplification and decay rates of the QNMs that are closely related to the SCC conjecture. Hence, they play an important role in our analysis. The surface gravities of RNdS black holes are at the inner horizon $\kappa_{\rm i}$, outer horizon $\kappa_{\rm o}$, and cosmological horizon $\kappa_{\rm c}$
\begin{align}\label{eq:kappa_ioc}
\kappa_{\rm i} &= \frac{(r_{\rm c}-r_{\rm i})(r_{\rm o}-r_{\rm i})(\rc+\ro+2\ri)}{2r_{\rm i}^2\left(\rc^2 + \ro^2 + (\rc+\ri)(\ro + \ri)\right)},\quad \kappa_{\rm o} = \frac{(r_{\rm c}-r_{\rm o})(r_{\rm o}-r_{\rm i})(\rc+2\ro+\ri)}{2r_{\rm o}^2\left(\rc^2 + \ro^2 + (\rc+\ri)(\ro + \ri)\right)},\\
\kappa_{\rm c} &= \frac{(r_{\rm c}-r_{\rm o})(r_{\rm c}-r_{\rm i})(2\rc+\ro+\ri)}{2r_{\rm c}^2\left(\rc^2 + \ro^2 + (\rc+\ri)(\ro + \ri)\right)},\quad 
\kappa_{\rm n}= \frac{(\rc + \ro + 2 \ri) (\rc + 2 \ro + \ri) (2 \rc + \ro + \ri)}{2(\rc + \ro + \ri)^{2}(\rc^2 + \ro^2 + (\rc+\ri)(\ro + \ri))},\nonumber
\end{align}
where $\kappa_{\rm n}$ is the surface gravity-like value at the negative solution of $\Delta(r)=0$: $r_\text{n}=-(r_\text{i}+r_\text{o}+r_\text{c})$. Hence, it has no physical meaning and is only used for simplifying equations. It is worth noting that there are two kinds of extremal black holes---one where the inner horizon coincides with the outer horizon, and one where the outer horizon coincides with the cosmological horizon. For convenience, the former is referred to as an extremal black hole in $\riro$ and the latter as an extremal black hole in $\rorc$.

\section{Effective Potential of Charged Particle}\label{sec3}

The unstable circular orbit of the particle in the eikonal limit is found to be associated with the decay rate of the QNM. Hence, we investigate the effective potential of the charged particle to demonstrate the unstable circular orbit in the spacetime of the RNdS black hole in Eq.\,\eqref{eq:metric02}. Here, the radial equation, including the effective potential, is obtained by the Hamilton--Jacobi method. The Hamiltonian is given as 
\begin{align}
{\cal H}&=\frac{1}{2}g^{\mu\nu}(p_\mu - qA_\mu)(p_\nu-qA_\nu), \label{eq:H}
\end{align}
in which the particle with the electric charge $q$ and four momenta $p_\mu$ is coupled with a gauge field. Then, the Hamilton--Jacobi action of the charged particle with the mass $m$ is written as
\begin{equation}\label{eq:S}
{\cal S}=\frac{1}{2}m^2 \tau - Et+{\cal S}_{r}(r)+{\cal S}_\theta\left(\theta\right)+L\phi,
\end{equation}
where the affine parameter denotes $\tau$. From the metric in Eq.\,(\ref{eq:metric02}), the translation symmetries can be read in time $t$ and azimuthal angle $\phi$. Then, the conserved quantities with respect to these translation symmetries are defined as the energy $E$ and angular momentum $L$ of the particle. According to the Hamilton--Jacobi method, the momenta of the particle is in terms of the action of Eq.\,(\ref{eq:S}) by ${\cal S}$ as $p_\mu=\partial_\mu S$. Then, substituting Eq.\,\eqref{eq:S} into Eq.\,\eqref{eq:H}, the Hamilton--Jacobi equation is obtained as
\begin{align}
-2 \frac{\partial {\cal S}}{\partial \tau}
&=-\frac{r^2}{\Delta}\left(-E+\frac{qQ}{r}\right)^2+\frac{\Delta}{r^2}(\partial_r {\cal S}_r(r))^2+\frac{1}{r^2}\left(\partial_\theta {\cal S}_\theta(\theta)\right)^2+\frac{L^2}{r^2\sin^2\theta} =m^2.
\end{align}
Here, the Hamilton--Jacobi equation can be separated into radial and $\theta$-directional equations by the separate constant ${\cal K}$ as
\begin{align}
{\cal K} &= -m^2 r^2+\frac{r^4}{\Delta}\left(-E+\frac{q Q}{r}\right)^2 - \Delta (\partial_r {\cal S}_r)^2,
\quad {\cal K} = (\partial_\theta {\cal S}_\theta)^2 +\frac{L^2}{ \sin^2\theta}. \label{eq:SrStheta}
\end{align}
Then, we can rewrite the Hamilton--Jacobi action as
\begin{align}\label{eq:I02}
{\cal S}=\frac{1}{2}m^2 \tau -Et +\int dr \sqrt{\rho(r)} +\int d\theta \sqrt{ \Theta(\theta)} +L\phi,
\end{align}
where the functions $\rho(r)$ and $\Theta(\theta)$ are given by
\begin{align}
\Sigma(r)=\frac{1}{\Delta}\left(-K-m^2 r^2 +\frac{r^4 }{\Delta}\left(-E+\frac{q Q}{r}\right)^2\right), \qquad \Xi(\theta)=K-\frac{1}{\sin^2\theta}L^2.
\end{align}
Then, full geodesic equations to the charged particle can be obtained from Eq.\,(\ref{eq:I02}) by the Hamilton--Jacobi method. Here, our concern is the unstable circular orbits on the equator; therefore, we can impose $\theta=\pi/2$ and $\dot{\theta}=0$ without the loss of generality, because the RNdS black hole has a spherical symmetric geometry. On the equator, the trajectory is obtained as 
\begin{align}\label{eq:dot_r}
 \dot{t} &= \frac{r^2}{\Delta}\left(E-\frac{qQ}{r}\right),\quad \dot{r} = \sqrt{\frac{\Delta}{r^2} \left(-m^2+\frac{r^2 }{\Delta}\left(-E+\frac{q Q}{r}\right)^2-\frac{1}{r^2}\frac{L^2}{\sin^2\theta} \right)},\quad \dot{\phi} = \frac{L}{r^2},
\end{align}
where a dot denotes a derivative with respect to the affine parameter $\tau$. Finally, the effective potential on the equator can be read from Eq.\,(\ref{eq:dot_r}) as
\begin{align}\label{eq:Veff1}
V_{\rm eff}\equiv \dot{r}^2= \frac{\Delta}{r^2} \left(-m^2+\frac{r^2 }{\Delta}\left(-E+\frac{q Q}{r}\right)^2-\frac{L^2}{r^2} \right), 
\end{align}
where $m^2>0$ for a timelike particle and $m^2=0$ for null particles. We can estimate possible orbits of the particle from the effective potential in Eq.\,(\ref{eq:Veff1}). The unstable circular orbit is obtained when the radius of the orbit coincides with the location of the maximum in the effective potential for the null case.

\section{Instability of Circular Orbits of Charged Null Particles}\label{sec4}

\begin{figure}[h]
  \begin{center}
\subfigure[{Metric function $\Delta(r)$}]{
\includegraphics[width=0.46\textwidth]{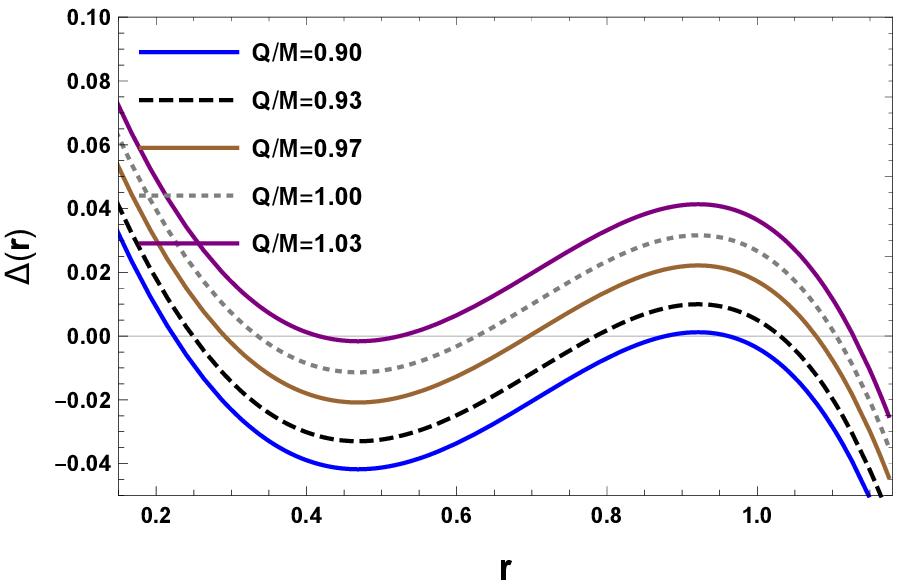}\label{metric2}} \,\,
\subfigure[{Instability time scale ${\cal T}_\lambda$}]{
\includegraphics[width=0.46\textwidth]{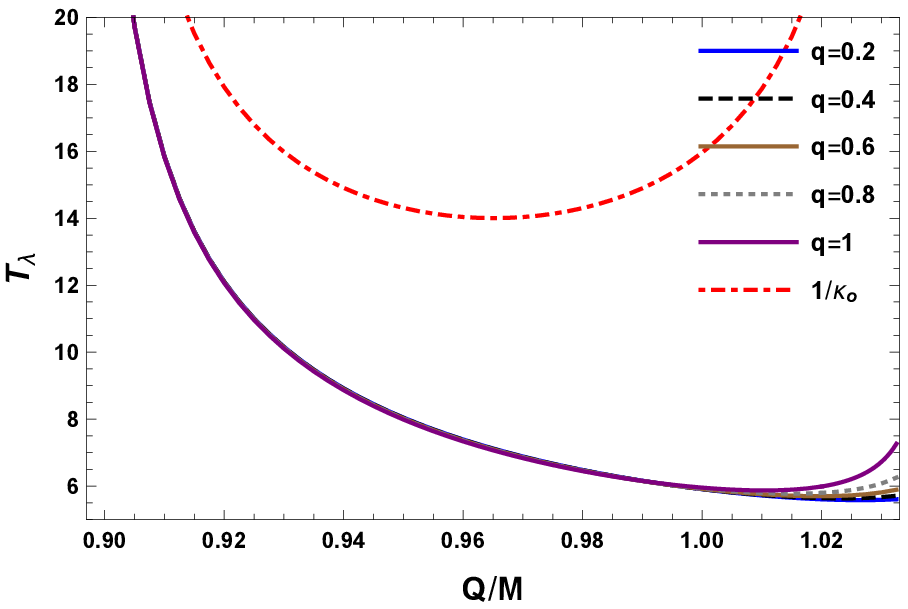}\label{fig:time_Lyapu1}} \\
\subfigure[{Typical orbital time scale ${\cal T}_\Omega$}]{
\includegraphics[width=0.46\textwidth]{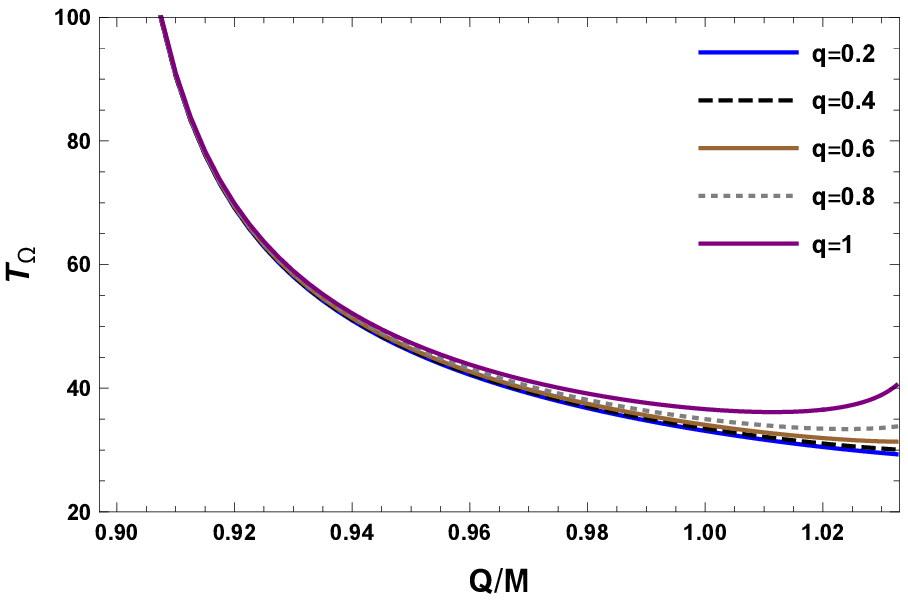}\label{fig:time_Omega1}} \,\,
\subfigure[{Critical exponent $\gamma$}]{
\includegraphics[width=0.46\textwidth]{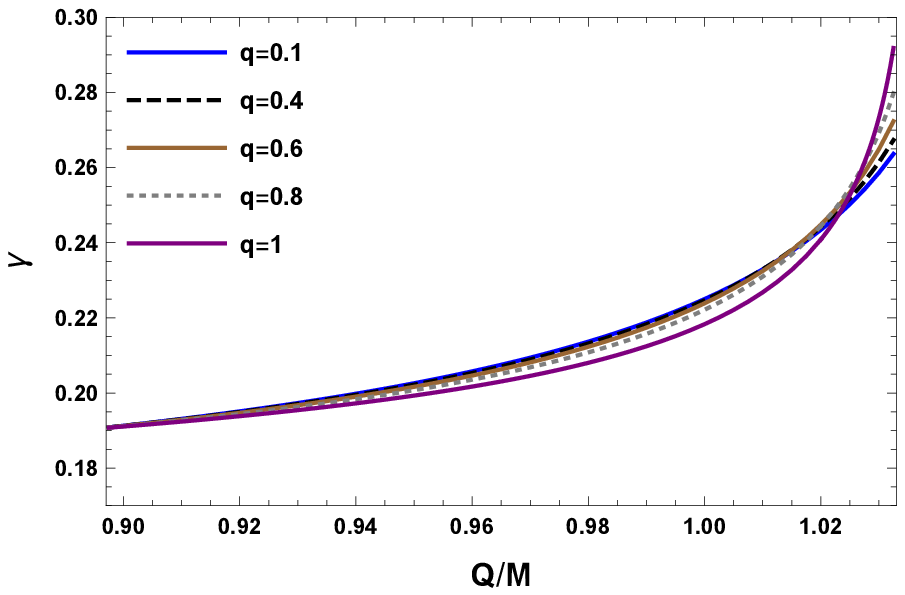}\label{fig:time_Omega_Lyapu1}}
  \end{center}
  \caption{Function $\Delta(r)$, instability time scale ${\cal T}_\lambda$, typical orbital time scale ${\cal T}_\Omega$, and critical exponent $\gamma$ with respect to the ratio of the charge to mass of an RNdS black hole for $E=1,\,\Lam=1,\,M=0.4$.}\label{fig:Lyapunov2}
\end{figure}

We can now estimate the motions of the particle from the effective potential in Eq.\,(\ref{eq:Veff1}). The unstable circular orbits of a null particle that is connected to the Lyapunov exponent are also obtained from the effective potential satisfying the expression
\begin{align}\label{eq:effective170a}
V_{\rm eff}(r_{\rm cir})=0,\quad V'_{\rm eff}(r_{\rm cir})=0, \quad V''_{\rm eff}(r_{\rm cir}) > 0,
\end{align} 
where the prime denotes derivatives with respect to the coordinate $r$. Furthermore,
\begin{align} \label{eq:DVeff1a}
V_{\rm eff}(\rcir)&=\frac{r_\text{cir}^2(-qQ+Er_\text{cir})^2-L^2\Delta(r_\text{cir})}{r_\text{cir}^4},\\
V'_{\rm eff}(\rcir)&=\frac{2qQ \rcir^2 (-q Q + E \rcir) + L^2 (4 \Delta(\rcir) - \rcir \Delta'(\rcir))}{\rcir^5},\nonumber\\
V''_{\rm eff}(r_{\rm cir }) &=\frac{2qQr_\text{cir}^2 (3qQ-2E r_\text{cir})-L^2 (20\Delta(r_\text{cir})+r_\text{cir}(-8\Delta'(r_\text{cir})+r_\text{cir}\Delta''(r_\text{cir}))}{r_\text{cir}^6},\nonumber
\end{align}
where
\begin{align}
\Delta'(\rcir)=-2 M + 2 \rcir - \frac{4 \Lambda}{3}\rcir^3, \quad
\Delta''(\rcir)=2 - 4\Lambda \rcir^2.
\end{align}
According to the first and second conditions in Eq.\,(\ref{eq:effective170a}) with $m=0$, the energy and angular momentum of the charged null particle are in the unstable circular orbits
\begin{align}\label{eq:Vprime}
& \frac{E}{q} =\frac{Q}{\rcir} \left(1+\frac{2\Delta(\rcir)}{\rcir\Delta'(\rcir)-4\Delta(\rcir)}\right), \qquad \frac{L}{q}= \frac{2Q\rcir\sqrt{\Delta(\rcir)}}{\rcir\Delta'(\rcir)-4\Delta(\rcir)},
\end{align}
Then, we can substitute the radius of the circular orbit $r_\text{cir}$ into the expressions for energy and angular momentum of the particle. The second derivative of the effective potential for the circular null orbits is thus rewritten as 
\begin{align} \label{eq:DDVeff}
V''_{\rm eff}(r_{\rm cir }) &=\frac{2 q^2 Q^2 \left(-8\Delta(\rcir)^2+\rcir^2\Delta'(\rcir)^2+4\rcir \Delta(\rcir)\Delta'(\rcir)-2\rcir^2 \Delta(\rcir)\Delta''(\rcir)\right)}{\rcir^4(\rcir\Delta'(\rcir)-4\Delta(\rcir))^2},
\end{align}
Furthermore, the first derivative of the coordinate time $t$ with respect to the affine parameter $\tau$ at the radius of the circular motion is rewritten as
\begin{align}
\left. \dot{t}~ \right|_{ r=\rcir}=\frac{2qQ\rcir}{\rcir \Delta'(\rcir)-4\Delta(\rcir)}. \label{eq:tdot} 
\end{align}
By combining Eq.\,(\ref{eq:DDVeff}) and Eq.\, (\ref{eq:tdot}), we obtain the Lyapunov exponent for the circular null geodesic \cite{Cardoso:2008bp} as follows
\begin{align}
\lambda_{\rm cir} &= \sqrt{\frac{V''_{\rm eff}(r_{\rm cir })}{2\dot{t}^2}}=\frac{1}{2\rcir^3}\sqrt{\rcir^2\Delta'(\rcir)^2-8\Delta(\rcir)^2+4\rcir \Delta(\rcir)\Delta'(\rcir)-2\rcir^2 \Delta(\rcir)\Delta''(\rcir)}.\label{eq:Lyapunov1} 
\end{align}
The Lyapunov exponent is closely related to the instabilities of the orbits. This can be shown in terms of the time scale named the instability time scale as ${\cal T}_\lambda\equiv 1/\lambda_{\rm cir}$. The instability time scale represents the time period for which the particle on the circular orbit leaks out of the trajectory. Since a particle on the unstable orbit cannot fill one period of the orbit entirely, the instability time is shorter than its typical orbital time ${\cal T}_\Omega\equiv2\pi/\Omega_{\rm cir}$, which is the time period of the orbit for one turn. Then, we can expect ${\cal T}_\lambda<{\cal T}_\Omega$, and this can be simply represented by the critical exponent \cite{Pretorius:2007jn} as
\begin{align}
\gamma = \frac{{\cal T}_\lambda}{{\cal T}_{\Omega}}.
\end{align}
If the critical exponent with respect to the inequality is $\gamma <1$, it implies that the particle leaks out of the circular null orbit before finishing a turn. Thus, the instability increases for a small critical exponent. For RNdS black holes, the instability and typical orbital time scales are obtained as
\begin{align} \label{eq:T_lambda}
{\cal T}_\lambda &=\frac{2\rcir^3}{\sqrt{\rcir^2\Delta'(\rcir)^2-8\Delta(\rcir)^2+4\rcir \Delta(\rcir)\Delta'(\rcir)-2\rcir^2 \Delta(\rcir)\Delta''(\rcir)} },\quad {\cal T}_\Omega = \frac{2\pi \rcir^2}{\sqrt{\Delta(\rcir)}},
\end{align}
where we use $\Omega_{\rm cir} = \sqrt{\Delta(\rcir)}/\rcir^2$. Then, the critical exponent is found to be
\begin{align}\label{eq:criticalexponent01a}
\gamma\equiv \frac{{\cal T}_\lambda}{{\cal T}_\Omega}=\frac{\rcir\sqrt{\Delta(\rcir)}}{\pi \sqrt{\rcir^2\Delta'(\rcir)^2-8\Delta(\rcir)^2+4\rcir \Delta(\rcir)\Delta'(\rcir)-2\rcir^2 \Delta(\rcir)\Delta''(\rcir)}}.
\end{align}
The instabilities of the circular orbits can be estimated from Eq.\,(\ref{eq:criticalexponent01a}) with respect to the radius $r_\text{cir}$. Note that there is a universal upper bound for the Lyapunov exponents of chaotic motions (i.e., $\lambda_{\rm cir} < \kappa_{\rm o}$), as discussed in \cite{Hashimoto:2016dfz}. This implies that the lower bound of the instability time scale is $\kappa_{\rm o}^{-1}<{\rm T}_\lambda$. However, the inequality is found to be violated for the charged particle in the RNdS black hole\cite{Zhao:2018wkl}. Thus, the instability time scale has no lower bound, which is consistent with our results. Since Eq.\,(\ref{eq:T_lambda}) and Eq.\,(\ref{eq:criticalexponent01a}) are given in terms of the radii of the unstable circular orbits, we depict their behaviors in the numerical analysis in Fig.\,\ref{fig:Lyapunov2}.
The states of RNdS black holes depend on the ratio of $Q$ and $M$, and the black holes only exist within specific ratios, as shown in Fig.\,\ref{metric2}. The minimum value of the ratio is $Q/M=0.9$, where the RNdS black hole is the extremal black hole in $r_\text{o}=r_\text{c}$. Then, as the ratio increases, the timelike spacetime becomes large. Finally, the ratio has the maximum value $Q/M=1.03$, when the black hole becomes the extremal black hole in $r_\text{i}=r_\text{o}$. Note that $\Lambda$ can be removed by rescaling the coordinates. This is effectively the same as setting $\Lambda=1$ in Fig.\,\ref{fig:Lyapunov2}. The instability and typical orbital time scales in Eq.\,(\ref{eq:T_lambda}) are represented in Fig.\,\ref{fig:time_Lyapu1} and Fig.\,\ref{fig:time_Omega1} with respect to the ratio $Q/M$. The instabilities of the orbits depend on $Q/M$ rather than $q$, and we can observe that $T_\lambda > T_\Omega$. Furthermore, the instability time scale is lower than the red dashed line in Fig.\,\ref{fig:time_Lyapu1}, and there is no lower bound on the instability time scale. This is consistent with the observations in \cite{Zhao:2018wkl}. The instability of the orbit increases as the ratio $Q/M$ increases. This can be observed in Fig.\,\ref{fig:time_Omega_Lyapu1} for the critical exponent. The critical exponent increases and is at a maximum at the extremal black hole in $r_\text{i}=r_\text{o}$. This implies that the orbit is most stable at the extremal black hole in $r_\text{i}=r_\text{o}$. Furthermore, the instability slightly decreases as the charge of the particle increases at the extremal black hole. 

\section{Relationship between Lyapunov Exponent and QNMs}\label{sec5}

The linearized relaxation of the scalar field is governed by the QNM in the spacetime of a black hole. Here, we prove that the Lyapunov exponent of the charged null particle is associated with the decay of the QNM under the charged scalar field in Nariai-type near-extremal RNdS black holes under a $r_\text{o}\approx r_\text{c}$, near Nariai-type extremal condition. Thus, from this point on, we focus on the Nariai-type near-extremal case.

\subsection{Decay Rate of Dominant QNM of Charged Scalar Field}\label{sec5.1}

We consider the massless scalar field with an electric charge in an RNdS black hole. Since the electric charge of the scalar field is coupled with the gauge field in the background, the covariant derivative should be introduced. Then, the field equation is given as
 \begin{equation}\label{eq:EOM}
\left((\nabla^\mu-iqA^\mu)(\nabla_\mu-iqA_\mu)-\mu^2\right)\Psi(t,r,\theta,\phi) =0,
\end{equation}
where $\mu$ and $q$ are the mass and charge of the scalar field, respectively. Since the field equation can be separated for each coordinate, the solution to the scalar field is written in the form
\begin{align}
\Psi(t,r,\theta,\phi)=\frac{\Phi(r)}{r} Y_{l m}(\theta)e^{-i\omega t} e^{im\phi},\label{eq:Psi}
\end{align}
where $\Phi(r)$ and $Y_{l m}(\theta)$ are the radial solution and spherical harmonics, respectively. The eigenvalues are $\omega$, $m$, and $l$ corresponding to the frequency and angular momenta of the scalar field, respectively. The radial equation is separately obtained according to Eqs.\,\eqref{eq:EOM} and \eqref{eq:Psi}. Furthermore, we assume the tortoise coordinate
\begin{equation}\label{eq:rtor}
r^*= \int \frac{r^2}{\Delta} dr=-\frac{\log(r_{\rm c}-r)}{2 \kappa_{\rm c}}+\frac{\log(r-r_{\rm o})}{2 \kappa_{\rm o}}-\frac{\log(r-r_{\rm i})}{2 \kappa_{\rm i}}+\frac{\log(r+\rc+\ro+\ri)}{2 \kappa_{\rm n}},
\end{equation}
where the range is $-\infty < r^* <\infty$. Then, the radial equation is simplified into the Schr\"{o}dinger-like equation as
\begin{equation}\label{eq:EOM_field}
\frac{d^2 \Phi}{d {r^*}^2}+\left(\left(\omega -\frac{qQ}{r}\right)^2 -{\cal V}_{\rm eff}(r)\right)\Phi =0,
\end{equation}
whose potential is
\begin{equation}\label{eq:V_eff_field}
{\cal V}_{\rm eff}(r) = \frac{\Delta(r)}{r^2}\left(\frac{l(l+1)}{r^2}+\mu^2+\frac{1}{r}\frac{d}{dr}\left(\frac{\Delta(r)}{r^2}\right)\right).
\end{equation}
At the outer and cosmological horizons, the solutions to the radial equation in Eq.\,(\ref{eq:EOM_field}) are obtained as
\begin{align}
\Phi(r^*) \sim e^{\pm i(\omega-qQ/\ro) r^*}, \quad r^*\to -\infty \, (r\to \ro); \quad \Phi(r^*) \sim e^{\pm i(\omega-qQ/\rc) r^*}, \quad r^*\to \infty \, (r\to \rc).
\end{align}
The QNMs in dS black holes are characterized to choose the boundary condition as\cite{Konoplya:2014lha} 
\begin{align}
\Phi(r^*) \sim \left\{\begin{array}{ll} e^{-i(\omega-qQ/\ro) r^*} \qquad & r^*\to -\infty \, (r\to \ro) \\ 
e^{i(\omega-qQ/\rc) r^*} \qquad & r^*\to \infty \,(r\to \rc) \end{array} \right. \label{eq:BD},
\end{align}
which implies that we only consider incoming flux of the scalar field at the outer horizon and outgoing flux at the cosmological horizon. For the boundary condition in Eq.\,(\ref{eq:BD}), we impose the Nariai-type near-extremal condition such that the outer horizon is located close to the cosmological horizon. This can be written as
\begin{equation}\label{eq:nearext}
\epsilon= \frac{\rc-\ro}{\ro} \ll 1. 
\end{equation}
We herein obtain the QNMs of the scalar field by the approximation to the P\"{o}schl--Teller potential, because the decay rate can be read in the imaginary part of the quasinormal frequency. In the approximation, the peak of the effective potential \eqref{eq:V_eff_field} has an important role. The location of the peak satisfies the condition
\begin{align}
\frac{d{\cal V}_{\rm eff}}{dr^*}\Big|_{r^*=r^*_{\rm p}}=0.
\end{align}
Furthermore, since we already assume the Nariai-type near-extremal condition, the peak has to be at the near-outer horizon; thus,
\begin{equation}\label{eq:nearext2a}
\delta = \frac{r_{\rm p}-r_{\rm o}}{r_{\rm o}}\ll 1.
\end{equation}
According to Eq.\,\eqref{eq:rtor} (and \ref{eq:nearext}), in the Nariai-type near-extremal limit, the location of the peak is approximately found in terms of the relationship between the radial and tortoise coordinates as
\begin{equation}\label{eq:rp}
r_{\rm p}= \frac{r_{\rm c}+r_{\rm o}e^{-2\kappa_{\rm o}r^*_{\rm p}}}{1+e^{-2\kappa_{\rm o}r^*_{\rm p}}}.
\end{equation}
Then, by Eq.~\eqref{eq:rp} and with our assumptions, the dimensionless parameters $\epsilon$ and $\delta$ are connected as
\begin{equation}\label{eq:}
\delta=\frac{1}{2} (1 + \tanh(\kappa_{\rm o}r^*_{\rm p}))\epsilon.
\end{equation}
Furthermore, the function $\Delta$ can be rewritten under the peak of the effective potential in Eq.\,(\ref{eq:V_eff_field}) as
\begin{equation}\label{eq:metric04}
\Delta(r_{\rm p}) = \frac{r_{\rm o}^2(r_{\rm o}-r_{\rm i})(r_{\rm i}+2r_{\rm o}+r_{\rm c})\Lambda \epsilon^2}{3} \frac{1}{4\cosh^2(\kappa_{\rm o}r^*_{\rm p})}.
\end{equation}
Hence, the effective potential is rewritten in the form of the P\"{o}schl--Teller potential at a Nariai-type near-extremal RNdS black hole as
\begin{align}\label{eq:PT}
{\cal V}_{\rm eff}&=\frac{{\cal V}_0}{\cosh^2(\kappa_{\rm o} r^*_{\rm p})},\quad {\cal V}_0=\frac{\Lambda \epsilon^2}{12}\left(\mu^2+\frac{l(l+1)}{r_{\rm o}^2}\right)(r_{\rm o}-r_{\rm i})(\ri+3\ro).
\end{align}
The quasinormal frequency in the P\"{o}schl--Teller potential can be generally read according to \cite{Ferrari:1984zz}. Therefore,
\begin{align}\label{eq:}
\omega &= \sqrt{{\cal V}_0-\frac{\kappa_{\rm o}^2}{4}} +\frac{qQ}{r_{\rm o}} -\frac{ q Q}{2\ro} \left(1+ \tanh\left(\kappa_{\rm o}r^*_{\rm p}\right)\right) \epsilon - i \left(n+\frac{1}{2}\right) \kappa_{\rm o},
\end{align}
where the real part of the frequency includes the coupling term between the charge and gauge field. Since the scalar field is associated with the unstable circular orbit of the particle in the eikonal limit, if we assume the condition of the eikonal limit, the quasinormal frequency is 
\begin{align}\label{eq:QNM}
\omega 
&= \frac{ l\epsilon}{2\ro}\sqrt{\frac{(\ro-\ri)(\ri+3\ro)}{3 \ro^2 + 2 \ro \ri + \ri^2}} - i \left(n+\frac{1}{2}\right) \kappa_{\rm o},
\end{align}
where the real and imaginary parts of the quasinormal frequency are given as
\begin{align}\label{eq:QNMrate1a}
\text{Re}(\omega)=\frac{ l\epsilon}{2\ro}\sqrt{\frac{(\ro-\ri)(\ri+3\ro)}{3 \ro^2 + 2 \ro \ri + \ri^2}},\quad \text{Im}(\omega)=-\left(n+\frac{1}{2}\right)\kappa_\text{o}.
\end{align} 
This is in agreement with the quasinormal mode in the eikonal limit in \cite{Hod:2018dpx}. Thus, the decay rate is obtained in the dominant QNM of the scalar field. Then,
\begin{align}\label{eq:QNMrate1b}
\text{Im}(\omega)=-\frac{1}{2}\kappa_\text{o}.
\end{align} 
Here, we expect that the decay rate is associated with the Lyapunov exponent of the charged null particle. Hence, the Lyapunov exponent should be rewritten under the same assumptions.

\subsection{Lyapunov Exponent in Near Nariai-Type Extremal Black Holes}\label{sec5.2}

We already obtained the decay rate of the dominant QNM of the massless scalar field in Eq.\,(\ref{eq:QNMrate1a}) for a Nariai-type near-extremal black hole in $r_\text{o}=r_\text{c}$. Furthermore, the Lyapunov exponent was also obtained for unstable circular orbits in an arbitrary state of an RNdS black hole. Here, we impose the same assumptions of the QNM to the Lyapunov exponent. Then, the Lyapunov exponent can be clearly related to the QNM. First, we assume the Nariai-type near-extremal limit in Eq.\,\eqref{eq:nearext}, which implies that the radius of the circular orbit is located in the near-horizon regime, $r_\text{cir}\approx r_\text{o}$. According to Eq.\,(\ref{eq:effective170a}) for the circular orbit, the radius is written in terms of $r_\text{o}$ and $\epsilon$.
\begin{align}
r_{\rm cir} &= r_{\rm o}+\frac{r_{\rm o}}{2}\epsilon + \mathcal{O}(\epsilon)^2.\label{eq:rcir}
\end{align}
Furthermore, for the radius of the circular orbit, the function $\Delta$ and its first derivative are rewritten as
\begin{align}\label{eq:}
 \Delta(\rcir)= \frac{\ro^2 (\ro - \ri) (3 \ro + \ri) \epsilon^2}{4 (3 \ro^2 + 2 \ro \ri + \ri^2)}, \quad
 \Delta'(\rcir)=\frac{\ro^3 \epsilon^2}{3 \ro^2 + 2 \ro \ri + \ri^2},
\end{align}
which are the second order of $\epsilon$, so their contributions are subtle. Then, the cosmological constant, mass, and electric charge of the black hole can be expressed in terms of the first order of $\epsilon$ owing to the Nariai-type near-extremal condition. From Eq.\,(\ref{eq:MQL}),
\begin{align}\label{eq:LQM2}
\Lam &= \frac{3}{3 \ro^2 + 2 \ro \ri + \ri^2} - \frac{ 9 \ro^2 +3 \ro \ri }{(3 \ro^2 + 2 \ro \ri + \ri^2)^2} \epsilon+\mathcal{O}(\epsilon^2),\\
 Q &= \frac{\ro \sqrt{\ri} \sqrt{2 \ro + \ri}}{\sqrt{3 \ro^2 + 2 \ro \ri + \ri^2}}+ \frac{ \ro \sqrt{\ri} (3 \ro + \ri) (\ro^2 + \ro \ri + \ri^2) }{2 \sqrt{2 \ro + \ri} (3 \ro^2 + 2 \ro \ri + \ri^2)^\frac{3}{2}}\epsilon+\mathcal{O}(\epsilon^2), \nonumber\\
M &= \frac{\ro (\ro + \ri)^2}{3 \ro^2 + 2 \ro \ri + \ri^2} +\frac{
 \ro (\ro + \ri) (3 \ro + \ri) (\ro^2 + \ri^2) \epsilon}{
 2 (3 \ro^2 + 2 \ro \ri + \ri^2)^2}+\mathcal{O}(\epsilon^2), \notag
\end{align}
where the charge is assumed to be positive for convenience. Furthermore,
the surface gravity on the outer horizon in Eq.~\eqref{eq:kappa_ioc} is also rewritten as 
\begin{equation}\label{eq:kappa_o_epsilon}
\kappa_{\rm o} = \frac{(r_{\rm o}-r_{\rm i})(3r_{\rm o}+r_{\rm i})\epsilon}{2r_{\rm o}(3r_{\rm o}^2+2r_{\rm o}r_{\rm i}+r_{\rm i}^2)}+\mathcal{O}(\epsilon^2), 
\end{equation} 
Then, in combination with Eqs.~\eqref{eq:rcir} and \eqref{eq:LQM2}, the angular velocity for the circular orbit on the Nariai-type near-extremal RNdS black hole is obtained as 
\begin{align}
\Omega_{\rm cir}= \frac{\epsilon}{2\ro }\sqrt{\frac{(\ro-\ri)(3\ro+\ri)}{3 \ro^2 + 2 \ro \ri + \ri^2}}. \label{eq:Omega_cir}
\end{align}
Moreover, the Lyapunov exponent in Eq.\,\eqref{eq:Lyapunov1} can be approximated as
\begin{align}
\lambda_{\rm cir}&= \frac{ (\ro - \ri) (3 \ro + \ri)\epsilon}{
 2 \ro (3 \ro^2 + 2 \ro \ri + \ri^2)}. \label{eq:Lyapunov_cir}
\end{align}
Then, the quasinormal frequency in Eq.\,(\ref{eq:QNMrate1a}) is clearly associated with the charged null particle in the first order of $\epsilon$. The quasinormal frequency is thus obtained as 
\begin{align}\label{eq:result01a}
\text{Re}(\omega)=\frac{ l\epsilon}{2\ro}\sqrt{\frac{(\ro-\ri)(\ri+3\ro)}{3 \ro^2 + 2 \ro \ri + \ri^2}}=l \Omega_\text{cir},\quad \text{Im}(\omega)=-\left(n+\frac{1}{2}\right)\kappa_\text{o}=-\left(n+\frac{1}{2}\right)\lambda_\text{cir}.
\end{align}
This implies that the QNMs of the massless scalar field with an electric charge corresponds to the unstable circular orbits of the charged null particle in the Nariai-type near-extremal RNdS black hole. In particular, the Lyapunov exponent is twice the decay rate of the dominant QNM in Eq.\,(\ref{eq:QNMrate1b}) as
\begin{align}\label{eq:result01b}
\lambda_\text{cir}=2\,\left|\text{Im}(\omega)\right|.
\end{align}
Therefore, the instabilities of the circular orbits for the charged null particle are proportional to the decay rates of the dominant QNMs for the charged massless scalar field.

\section{Strong Cosmic Censorship Conjecture}\label{sec6}

Here, we investigated the validity of the SCC conjecture in Nariai-type near-extremal RNdS black holes by the Lyapunov exponent of the charged null particle. According to Eq.\,(\ref{eq:result01a}) and Eq.\,(\ref{eq:result01b}), the Lyapunov exponent of the charged null particle corresponds to the decay rate of the dominant QNM of the charged massless scalar field. In the SCC conjecture, destabilizing the Cauchy horizon plays a significant role, and destabilization depends on the decay rate of the dominant QNM because the decay rate determines whether the dominant QNM is infinitely blue-shifted or not. Since the blueshift occurs owing to the Cauchy horizon, the amplification rate is fixed to the surface gravity of the inner horizon. On the contrary, the decay occurs exponentially as 
\begin{equation}\label{eq:decay1}
|\Psi-\Psi_0|\sim e^{-\eta t},
\end{equation}
where the spectral gap is $\eta$ \cite{Barreto:1997, Bony:2008, Dyatlov:2012, Dyatlov:2015}. Hence, we have to find the decay rate of the QNMs to determine its amplification or decay. At the Cauchy horizon, the divergence of the scalar field energy is related to the ratio $\beta\equiv \eta/\kappa_\text{i}$\cite{Dias:2018ufh} in the SCC conjecture given by \cite{Christodoulou:2008nj}. The spectral gap is herein equivalent to $\left| \text{Im}(\omega) \right|$. When $\beta>\frac{1}{2}$, the decay of the charged massless scalar field is dominant, and the singularity is timelike because the Cauchy horizon is still stable. In this case, the SCC conjecture is violated. However, if $\beta<\frac{1}{2}$, the Cauchy horizon can be destabilized by the infinitely blue-shifted scalar field. Thus, the SCC conjecture is valid. In our analysis, we observe that the decay rate of the dominant QNM is half that of the Lyapunov exponent of the charged null particle in the Nariai-type near-extremal black hole. Hence,
\begin{align}
\frac{\lambda_\text{cir}}{2}=\left| \text{Im}(\omega) \right|=\frac{\kappa_\text{o}}{2}<\frac{|\kappa_\text{i}|}{2}.
\end{align}
This implies that the amplification of the charged massless scalar field is dominant in Nariai-type near-extremal black holes, and $\beta<1/2$. Therefore, the SCC conjecture is valid in our analysis based on the compact relationships in Eq.\,(\ref{eq:result01a}) and Eq.\,(\ref{eq:result01b}). This is consistent with the analysis in QNMs\cite{Mo:2018nnu, Hod:2018dpx}. Our compact relationship is valid in the first order of $\epsilon$, so our conclusion about the SCC conjecture is also based on the first order. Note that, in the analysis of QNMs of the charged scalar field, non-perturbative effects of the electric charge become important in the large $q$ expansion. Owing to the non-perturbative effects, the SCC conjecture can be violated under the charged scalar field\cite{Dias:2018ufh}.

\section{Summary}\label{sec7}

We investigated the validity of the SCC conjecture in Nariai-type near-extremal RNdS black holes using the Lyapunov exponent of the charged null particle in a circular null orbit. Since the SCC conjecture has to be discussed in QNMs, the relationship between the Lyapunov exponent and QNMs is generalized to the case of the charged massless scalar field corresponding to the charged null particle. Then, we compute the Lyapunov exponent representing the instabilities of orbits in RNdS black holes. By using the instability time scale associated with the Lyapunov exponent, the circular null orbits of the charged null particle around the extremal black hole in $\riro$ are more stable than those around the extremal black hole in $\rorc$. Furthermore, since the decay rate in the QNM with the eikonal limit can be written in the analytical form in the case of a Nariai-type near-extremal RNdS black hole, we imposed the same assumption on the Lyapunov exponent of the charged null particle. Thus, we found the compact relationship between the Lyapunov exponent and QNMs in Eq.\,(\ref{eq:result01a}) for the charged case. According to the compact relationship in the charged case, the decay rate in a Nariai-type near-extremal black hole of $\rorc$ represents $\beta<1/2$. This implies that the SCC conjecture is valid in our analysis based on the Lyapunov exponent.

\vspace{10pt} 

\noindent{\bf Acknowledgments}

\noindent This work was supported by the National Research Foundation of Korea (NRF) grant funded by the Korea government (MSIT) (NRF-2018R1C1B6004349) and the Dongguk University Research Fund.


\begin{thebibliography}{99}

\bibitem{Hawking:1974sw} 
  S.~W.~Hawking,
  Commun.\ Math.\ Phys.\  {\bf 43}, 199 (1975).

\bibitem{Hawking:1976de} 
  S.~W.~Hawking,
  Phys.\ Rev.\ D {\bf 13}, 191 (1976).

\bibitem{Bekenstein:1973ur} 
  J.~D.~Bekenstein,
  Phys.\ Rev.\ D {\bf 7}, 2333 (1973).

\bibitem{Bekenstein:1974ax} 
  J.~D.~Bekenstein,
  Phys.\ Rev.\ D {\bf 9}, 3292 (1974).

\bibitem{Penrose:1964wq} 
  R.~Penrose,
  Phys.\ Rev.\ Lett.\  {\bf 14}, 57 (1965).

\bibitem{Penrose:1969pc} 
  R.~Penrose,
  Riv.\ Nuovo Cim.\  {\bf 1}, 252 (1969)
  [Gen.\ Rel.\ Grav.\  {\bf 34}, 1141 (2002)].

\bibitem{Hawking:1969sw} 
  S.~W.~Hawking and R.~Penrose,
  Proc.\ Roy.\ Soc.\ Lond.\ A {\bf 314}, 529 (1970).

\bibitem{Wald:1974ge}
 R.~Wald,
 Annals\ Phys.\ {\bf 82}, 548 (1974).

\bibitem{Jacobson:2009kt} 
  T.~Jacobson and T.~P.~Sotiriou,
  Phys.\ Rev.\ Lett.\  {\bf 103}, 141101 (2009).

\bibitem{Barausse:2010ka} 
  E.~Barausse, V.~Cardoso and G.~Khanna,
  Phys.\ Rev.\ Lett.\  {\bf 105}, 261102 (2010).

\bibitem{Barausse:2011vx} 
  E.~Barausse, V.~Cardoso and G.~Khanna,
  Phys.\ Rev.\ D {\bf 84}, 104006 (2011).

\bibitem{Colleoni:2015ena} 
  M.~Colleoni, L.~Barack, A.~G.~Shah and M.~van de Meent,
  Phys.\ Rev.\ D {\bf 92}, no. 8, 084044 (2015).

\bibitem{Colleoni:2015afa} 
  M.~Colleoni and L.~Barack,
  Phys.\ Rev.\ D {\bf 91}, 104024 (2015).

\bibitem{Sorce:2017dst} 
  J.~Sorce and R.~M.~Wald,
  Phys.\ Rev.\ D {\bf 96}, no. 10, 104014 (2017).

\bibitem{Hubeny:1998ga} 
  V.~E.~Hubeny,
  Phys.\ Rev.\ D {\bf 59}, 064013 (1999).

\bibitem{Isoyama:2011ea} 
  S.~Isoyama, N.~Sago and T.~Tanaka,
  Phys.\ Rev.\ D {\bf 84}, 124024 (2011).

\bibitem{Crisostomo:2003xz} 
  J.~Crisostomo and R.~Olea,
  Phys.\ Rev.\ D {\bf 69}, 104023 (2004).

\bibitem{Gao:2012ca} 
  S.~Gao and Y.~Zhang,
  Phys.\ Rev.\ D {\bf 87}, no. 4, 044028 (2013).

\bibitem{Hod:2013vj} 
  S.~Hod,
  Phys.\ Rev.\ D {\bf 87}, no. 2, 024037 (2013).

\bibitem{Rocha:2014jma}
  J.~V.~Rocha and R.~Santarelli,
  Phys.\ Rev.\ D {\bf 89}, no. 6, 064065 (2014).

\bibitem{Gwak:2015fsa} 
  B.~Gwak and B.~H.~Lee,
  JCAP {\bf 1602}, 015 (2016).

\bibitem{Cardoso:2015xtj} 
  V.~Cardoso and L.~Queimada,
  Gen.\ Rel.\ Grav.\  {\bf 47}, no. 12, 150 (2015).

\bibitem{Horowitz:2016ezu} 
  G.~T.~Horowitz, J.~E.~Santos and B.~Way,
  Class.\ Quant.\ Grav.\  {\bf 33}, no. 19, 195007 (2016).

\bibitem{Revelar:2017sem} 
  K.~S.~Revelar and I.~Vega,
  Phys.\ Rev.\ D {\bf 96}, no. 6, 064010 (2017).

\bibitem{Gwak:2017kkt} 
  B.~Gwak,
  JHEP {\bf 1711}, 129 (2017).

\bibitem{Duztas:2017lxk} 
  K.~D\"{u}zta\c{s},
  Class.\ Quant.\ Grav.\  {\bf 35}, no. 4, 045008 (2018).

\bibitem{Ge:2017vun} 
  B.~Ge, Y.~Mo, S.~Zhao and J.~Zheng,
  Phys.\ Lett.\ B {\bf 783}, 440 (2018).

\bibitem{Yu:2018eqq} 
  T.~Y.~Yu and W.~Y.~Wen,
  Phys.\ Lett.\ B {\bf 781}, 713 (2018).

\bibitem{Gim:2018axz} 
  Y.~Gim and B.~Gwak,
  Phys.\ Lett.\ B {\bf 794}, 122 (2019).

\bibitem{Shaymatov:2018fmp} 
  S.~Shaymatov, N.~Dadhich and B.~Ahmedov,
  Eur.\ Phys.\ J.\ C {\bf 79}, no. 7, 585 (2019).

\bibitem{Zeng:2019jrh} 
  X.~X.~Zeng, Y.~W.~Han and D.~Y.~Chen,
  arXiv:1901.08915 [gr-qc].

\bibitem{Semiz:2005gs} 
  I.~Semiz,
  Gen.\ Rel.\ Grav.\  {\bf 43}, 833 (2011).

\bibitem{Hod:2008zza} 
  S.~Hod,
  Phys.\ Rev.\ Lett.\  {\bf 100}, 121101 (2008).

\bibitem{Toth:2011ab} 
  G.~Z.~Toth,
  Gen.\ Rel.\ Grav.\  {\bf 44}, 2019 (2012).

\bibitem{Duztas:2013wua} 
    K.~D\"{u}zta\c{s}, and \.{I}.~Semiz,
  Phys.\ Rev.\ D {\bf 88}, no. 6, 064043 (2013).

\bibitem{Semiz:2015pna} 
  \.{I}.~Semiz and   K.~D\"{u}zta\c{s},
  Phys.\ Rev.\ D {\bf 92}, no. 10, 104021 (2015).

\bibitem{Natario:2016bay} 
  J.~Natario, L.~Queimada and R.~Vicente,
  Class.\ Quant.\ Grav.\  {\bf 33}, no. 17, 175002 (2016).

\bibitem{Gwak:2018akg} 
  B.~Gwak,
  JHEP {\bf 1809}, 081 (2018).

\bibitem{Duztas:2018adf} 
  K.~D\"{u}zta\c{s},
  Int.\ J.\ Mod.\ Phys.\ D {\bf 28}, no. 02, 1950044 (2019).

\bibitem{Chen:2018yah} 
  D.~Chen,
  arXiv:1812.03459 [gr-qc].

\bibitem{Chen:2019nsr} 
  D.~Chen, W.~Yang and X.~Zeng,
  Nucl.\ Phys.\ B {\bf 946}, 114722 (2019).

\bibitem{Gwak:2019asi} 
  B.~Gwak,
  JCAP {\bf 1908}, 016 (2019).

\bibitem{Gwak:2019rcz} 
  B.~Gwak,
  arXiv:1910.13329 [gr-qc].

\bibitem{Hawking:1970} 
  S.~W.~Hawking and R.~Penrose,
  Proc.\ R.\ Soc.\ London\ A\ {\bf 314}, 529 (1970).

\bibitem{Chambers:1997ef} 
  C.~M.~Chambers,
  Annals Israel Phys.\ Soc.\  {\bf 13}, 33 (1997).

\bibitem{Matzner:1979zz} 
  R.~A.~Matzner, N.~Zamorano and V.~D.~Sandberg,
  Phys.\ Rev.\ D {\bf 19}, 2821 (1979).

\bibitem{Poisson:1990eh} 
  E.~Poisson and W.~Israel,
  Phys.\ Rev.\ D {\bf 41}, 1796 (1990).

\bibitem{Ori:1991zz} 
  A.~Ori,
  Phys.\ Rev.\ Lett.\  {\bf 67}, 789 (1991).

\bibitem{Brady:1995ni} 
  P.~R.~Brady and J.~D.~Smith,
  Phys.\ Rev.\ Lett.\  {\bf 75}, 1256 (1995).

\bibitem{Brady:1998au} 
  P.~R.~Brady, I.~G.~Moss and R.~C.~Myers,
  Phys.\ Rev.\ Lett.\  {\bf 80}, 3432 (1998).
  
  \bibitem{Cardoso:2017soq} 
  V.~Cardoso, J.~L.~Costa, K.~Destounis, P.~Hintz and A.~Jansen,
  Phys.\ Rev.\ Lett.\  {\bf 120}, no. 3, 031103 (2018).
  
\bibitem{Luna:2018jfk} 
  R.~Luna, M.~Zilhao, V.~Cardoso, J.~L.~Costa and J.~Natario,
  Phys.\ Rev.\ D {\bf 99}, no. 6, 064014 (2019).

\bibitem{Destounis:2018qnb} 
  K.~Destounis,
  Phys.\ Lett.\ B {\bf 795}, 211 (2019).

\bibitem{Rahman:2018oso} 
  M.~Rahman, S.~Chakraborty, S.~SenGupta and A.~A.~Sen,
  JHEP {\bf 1903}, 178 (2019).

  \bibitem{Ge:2018vjq} 
  B.~Ge, J.~Jiang, B.~Wang, H.~Zhang and Z.~Zhong,
  JHEP {\bf 1901}, 123 (2019).

\bibitem{Gwak:2018rba} 
  B.~Gwak,
  Eur.\ Phys.\ J.\ C {\bf 79}, no. 9, 767 (2019).

\bibitem{Burikham:2017gdm} 
  P.~Burikham, S.~Ponglertsakul and L.~Tannukij,
  Phys.\ Rev.\ D {\bf 96}, no. 12, 124001 (2017).

\bibitem{Hod:2018dpx} 
  S.~Hod,
  Nucl.\ Phys.\ B {\bf 941}, 636 (2019).
  
  \bibitem{Cardoso:2018nvb} 
  V.~Cardoso, J.~L.~Costa, K.~Destounis, P.~Hintz and A.~Jansen,
  Phys.\ Rev.\ D {\bf 98}, no. 10, 104007 (2018).
  
  \bibitem{Dias:2018etb} 
  O.~J.~C.~Dias, H.~S.~Reall and J.~E.~Santos,
  JHEP {\bf 1810}, 001 (2018).
  
  \bibitem{Mo:2018nnu} 
  Y.~Mo, Y.~Tian, B.~Wang, H.~Zhang and Z.~Zhong,
  Phys.\ Rev.\ D {\bf 98}, no. 12, 124025 (2018).
  
  \bibitem{Dias:2018ufh} 
  O.~J.~C.~Dias, H.~S.~Reall and J.~E.~Santos,
  Class.\ Quant.\ Grav.\  {\bf 36}, no. 4, 045005 (2019).

\bibitem{Liu:2019rbq} 
  X.~Liu, S.~Van Vooren, H.~Zhang and Z.~Zhong,
  JHEP {\bf 1910}, 186 (2019).

\bibitem{Zhang:2019ynp} 
  M.~Zhang, J.~Jiang and Z.~Zhong,
  Physics Letters B, Volume 798 (2019) 134959.

\bibitem{Hod:2018lmi} 
  S.~Hod,
  Phys.\ Lett.\ B {\bf 780}, 221 (2018).

\bibitem{Liu:2019lon} 
  H.~Liu, Z.~Tang, K.~Destounis, B.~Wang, E.~Papantonopoulos and H.~Zhang,
  JHEP {\bf 1903}, 187 (2019).

\bibitem{Gwak:2019ttv} 
  B.~Gwak,
  arXiv:1903.11758 [gr-qc].

\bibitem{Guo:2019tjy} 
  H.~Guo, H.~Liu, X.~M.~Kuang and B.~Wang,
  Eur.\ Phys.\ J.\ C {\bf 79}, no. 11, 891 (2019).

\bibitem{Gan:2019jac} 
  Q.~Gan, G.~Guo, P.~Wang and H.~Wu,
  arXiv:1907.04466 [hep-th].

\bibitem{Destounis:2019omd} 
  K.~Destounis, R.~D.~B.~Fontana, F.~C.~Mena and E.~Papantonopoulos,
  JHEP {\bf 1910}, 280 (2019).

  \bibitem{Mashhoon:1985cya} 
  B.~Mashhoon,
  Phys.\ Rev.\ D {\bf 31}, no. 2, 290 (1985).

  \bibitem{Cardoso:2008bp} 
  V.~Cardoso, A.~S.~Miranda, E.~Berti, H.~Witek and V.~T.~Zanchin,
  Phys.\ Rev.\ D {\bf 79}, 064016 (2009).

\bibitem{Dias:2018ynt} 
  O.~J.~C.~Dias, F.~C.~Eperon, H.~S.~Reall and J.~E.~Santos,
  Phys.\ Rev.\ D {\bf 97}, no. 10, 104060 (2018).

\bibitem{Pretorius:2007jn} 
  F.~Pretorius and D.~Khurana,
  Class.\ Quant.\ Grav.\  {\bf 24}, S83 (2007).

\bibitem{Hashimoto:2016dfz} 
  K.~Hashimoto and N.~Tanahashi,
  Phys.\ Rev.\ D {\bf 95}, no. 2, 024007 (2017).

\bibitem{Zhao:2018wkl} 
  Q.~Q.~Zhao, Y.~Z.~Li and H.~Lu,
  Phys.\ Rev.\ D {\bf 98}, no. 12, 124001 (2018).

\bibitem{Konoplya:2014lha} 
  R.~A.~Konoplya and A.~Zhidenko,
  Phys.\ Rev.\ D {\bf 90}, no. 6, 064048 (2014).

\bibitem{Ferrari:1984zz} 
  V.~Ferrari and B.~Mashhoon,
  Phys.\ Rev.\ D {\bf 30}, 295 (1984).

\bibitem{Barreto:1997} 
 A. S\'{a} Barreto and M. Zworski,
  Math. Res. Lett. 4, 103 (1997).
  
 \bibitem{Bony:2008} 
 J.-F. Bony and D. H\"{a}fner,
 Commun. Math. Phys. 282, 697 (2008).
  
\bibitem{Dyatlov:2012} 
 S. Dyatlov,
 Ann. Henri Poincar\'{e} 13, 1101 (2012).
 
\bibitem{Dyatlov:2015} 
 S. Dyatlov,
Commun. Math. Phys. 335, 1445 (2015).

\bibitem{Christodoulou:2008nj} 
  D.~Christodoulou,
  arXiv:0805.3880 [gr-qc].



\end{thebibliography}
\end{document}